\documentclass[utf8]{frontiersFPHY} % for Physics and Applied Mathematics and Statistics articles

\usepackage{url,hyperref,lineno,microtype,subcaption,float}
\usepackage[onehalfspacing]{setspace}
\def\keyFont{\fontsize{8}{11}\helveticabold }
\def\firstAuthorLast{Srivastava {et~al.}} %use et al only if is more than 1 author
\def\Authors{A.K.~Srivastava\,$^{1,*}$, S.~W.~McIntosh\,$^{2}$, N.~Arge\,$^{3}$, D.~Banerjee\,$^{4}$,
E.~Cliver\,$^{5}$, M.~Dikpati\,$^{2}$, B.N.~Dwivedi\,$^{1}$, M. Guhathakurta\,$^{6}$, B.B.~Karak\,$^{1,2}$,
R.J.~Leamon\,$^{3,7}$, P.~Martens\,$^{8}$, S.K.~Matthew\,$^{9}$, A.~Munoz-Jaramillo\,$^{9}$, D.~Nandi\,$^{10}$,
A.~Norton\,$^{11}$, L.~Upton\,$^{2}$, S.~Chatterjee\,$^{4}$, R.~Mazumder\,$^{10}$, Yamini K.~Rao\,$^{1}$, and R.~Yadav\,$^{9}$}
\def\Address{$^{1}$ Department of Physics,  Indian Institute of Technology (BHU), Varanasi-221005, India.\\
$^{2}$ High Altitude Observatory, National Center for Atmospheric Research, PO Box 3000, Boulder, Colorado 80307, USA. \\
$^{3}$ Heliophysics Division, NASA Goddard Space Flight Center, Greenbelt, Maryland 20771, USA. \\
$^{4}$ Indian Institute of Astrophysics, Koramangala, Bangalore 560034, India. \\
$^{5}$ National Solar Observatory, Boulder, CO, USA. \\
$^{6}$ NASA Headquarters, USA. \\
$^{7}$ Department of Astronomy, University of Maryland College Park, Maryland 20742, USA. \\
$^{8}$ Department of Physics and Astronomy, Georgia State University, Atlanta, GA 30303, USA. \\
$^{9}$ Udaipur Solar Observatory, Physical Research Laboratory, Badi Road, Udaipur 313001, India. \\
$^{10}$ Center of Excellence in Space Sciences India, IISER Kolkata, Mohanpur 741246, West Bengal, India. \\
$^{11}$ W.W. Hansen Experimental Physics Laboratory, Stanford University, USA. }
\def\corrAuthor{Dr. A.K. Srivastava}

\def\corrEmail{asrivastava.app@iitbhu.ac.in}

\begin{document}
\onecolumn
\firstpage{1}

%\title[Extended Solar Cycles]{Extended Solar Cycles (ESCs) : A  Tool to Understanding the Sun's Magnetism at Diverse Spatio-temporal Scales} 

\title[Extended Solar Cycles]{The Extended Solar Cycle: Muddying the Waters of Solar/Stellar Dynamo Modeling Or Providing Crucial Observational Constraints?} 

\author[\firstAuthorLast ]{\Authors} %This field will be automatically populated
\address{} %This field will be automatically populated
\correspondance{} %This field will be automatically populated

\extraAuth{}% If there are more than 1 corresponding author, comment this line and uncomment the next one.
%\extraAuth{corresponding Author2 \\ Laboratory X2, Institute X2, Department X2, Organization X2, Street X2, City X2 , State XX2 (only USA, Canada and Australia), Zip Code2, X2 Country X2, email2@uni2.edu}

\maketitle

\begin{abstract}
In 1844 Schwabe discovered that the number of sunspots increased and decreased over a period of about 11 years, that variation became known as the sunspot cycle. Almost eighty years later, Hale described the nature of the Sun's magnetic field, identifying that it takes about 22 years for the Sun's magnetic polarity to cycle. It was also identified that the latitudinal distribution of sunspots resembles the wings of a butterfly showing migration of sunspots in each hemisphere that abruptly start at mid-latitudes towards the Sun's equator over the next 11 years. These sunspot patterns were shown to be asymmetric across the equator. In intervening years, it was deduced that the Sun (and sun-like stars) possess magnetic activity cycles that are assumed to be the physical manifestation of a dynamo process that results from complex circulatory transport processes in the star's interior. Understanding the Sun's magnetism, its origin and its variation, has become a fundamental scientific objective \-- the distribution of magnetism, and its interaction with convective processes, drives various plasma processes in the outer atmosphere. In the past few decades, a range of diagnostic techniques have been employed to systematically study finer scale magnetized objects, and associated phenomena. The patterns discerned became known as the ``Extended Solar Cycle'' (ESC). The patterns of the ESC appeared to extend the wings of the activity butterfly back in time, nearly a decade before the formation of the sunspot pattern, and to much higher solar latitudes. In this short review, we describe their observational patterns of the ESC and discuss possible connections to the solar dynamo as we depart on a multi-national collaboration to investigate the origins of solar magnetism through a blend of archived and contemporary data analysis with the goal of improving solar dynamo understanding and modeling.   
\tiny
 \keyFont{ \section{Keywords:} Sun: Magnetism, Sun: Interior, Sun: Rotation, Solar Cycle, Sunspots} %All article types: you may provide up to 8 keywords; at least 5 are mandatory.
\end{abstract}

%\section{Introduction}
%
The Sun is a magnetically active star. It possesses a magnetic field of complex evolving geometry that extends far out towards interplanetary space after its origin in the Sun's interior (e.g., Fan, 2009). The magnetic field of the Sun, structured in space and time over disparate scales is maintained by a dynamo that operates within the convecting plasma confined roughly to the outer 30\% of the Sun's radius. The net output of that dynamo waxes and wanes in strength every 11 years (see, e.g., Usoskin, 2017). Further, understanding the mechanism, or mechanisms, by which the magnetic field traps the Sun's sub-surface energy reservoir to couple the subsurface layers with those of the outer atmosphere to transport, build-up and dissipate colossal amounts of energy there poses a perennial challenge. In other words, the persistent circulative and convective forcing of that magnetic field drives radiative and particulate activity across time-scales, including what we now call ``space weather'' (e.g., Svalgaard, 2013). The reliance of society on space-based technology has reached a point where understanding the origin, evolution, and multi-faceted impacts of the Sun's magnetism has reached a heightened level of importance in astrophysics. Beyond this very practical issue, there is an implicit understanding that unlocking the puzzle around the origins of the Sun's magnetism could provide vital clues to the origins and behavior of the magnetic field in other stars (see, e.g., Lanza, 2010). 

\begin{figure}[!t]
\begin{center}
\includegraphics[width=15cm]{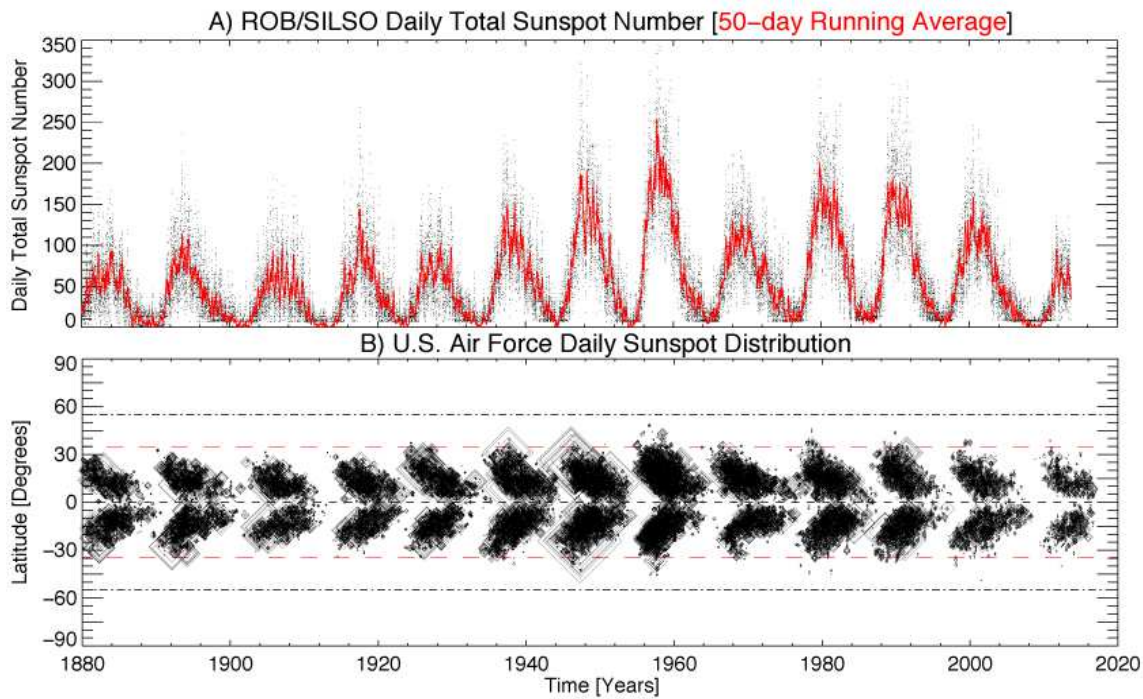}
\end{center}
\caption{The observed variation in sunspot number (top) and latitude (bottom). In panel A, we show the daily total sunspot number with a 50-day running average overplotted in red. The $\sim$11-year quasi-cyclic variability is clear. In panel B, the sunspot locations are shown, their extent is indicated by the size of the symbol. Note that the sunspot butterfly wings grow abruptly at latitudes largely inside $\sim$35$^{\circ}$. For reference we show red dashed and black dot\--dashed lines at 35$^{\circ}$ and 55$^{\circ}$ latitude.}\label{f1}
\end{figure}

The variation in number of the canonical marker of solar magnetic activity, sunspots, would appear to demonstrate that the Sun's magnetic field is routinely circulated with a periodicity close to 11 years (e.g., Hathaway, 2010). The gross variation in sunspot number, and variation of the sunspot distribution in latitude exemplify this quasi\--periodic variation, see e.g., Fig.~\ref{f1}. There we see that, with every new sunspot cycle, a fresh set of spots takes over the Sun's surface in a peculiar pattern that starts abruptly at mid solar latitudes and slowly migrates towards the equator (over a span of around 9 years) that takes on the shape of a butterfly's wing (Maunder, 1904). In this ``butterfly plot'' the wing like pattern of sunspot migration repeats again, and again. The determination that sunspots were locations of concentrated (strong) magnetic field (Hale, 1908; Hale et al., 1919) and the realization that, in each hemisphere of the Sun, the butterfly's wings alternated in magnetic polarity led to the stunning realization that the Sun's magnetic field was reversing approximately every 22-years (Hale and Nicholson, 1925).

Many features of cyclic solar variability have been cataloged extensively through decades of observation, e.g., cyclic variation of sunspot number approximately every 11 years, polarity reversal of the magnetic field in 22 years (Usoskin, 2017), the slight north-south asymmetry in activity where the sunspot production of the hemispheres will lead or lag by upto a couple of years  (e.g., Vizoso and Ballester, 1990; McIntosh et al., 2013), emergence of the spot groups with east-west magnetic polarity in each hemisphere (known as “Hale’s Law”; Hale et al., 1919), the emergence of active region (or sunspot pairs) with one magnetic polarity slightly more northward than the other (known as “Joy’s law”; Hale et al.,
1919). These observations have motivated extensive investigations and modeling efforts (too many to reference here) to explain the production of these magnetic patterns via the convective circulation of the solar interior. The most prominently recognized efforts, the dynamo wave model (e.g., Parker, 1955; Yoshimura, 1975), the “Babcock-Leighton” model (e.g., Babcock, 1961; Leighton, 1969), the ``Babcock\--Leighton'' model (e.g., Babcock, 1961; Leighton, 1969), and the ``Flux Transport Dynamo'' (e.g., Wang and Sheeley, 1991). These models, and the underlying theory, based {\em solely} on the magneto-spatio-temporal progression of sunspots, is expertly summarized by Charbonneau (2010).

These magnetohydrodynamic (MHD) dynamo models have been considered, for the last few decades, as the most viable explanation for the process responsible for the generation, and sustenance of the 22-year magnetic activity cycle, the 11-year solar cycle, and the appearance of the butterfly diagram. At very highest level, these dynamo processes capture: (i) generation of toroidal field in the solar tachocline by differential rotation, (ii) generation of poloidal field at the surface as per the Babcock-Leighton process, (iii) advection by meridional circulation, and (iv) cycle irregularities by the fluctuations imposed in the Babcock-Leighton process (e.g., Choudhuri et al., 1995, 2007; Dikpati and Gilman, 2006;
Jiang et al., 2007; Yeates et al., 2008, and references cited therein). The broadly acknowledged difficulty in forecasting/predicting sunspot cycle timing and amplitude has led some to speculate that the sunspot number and patterns do not form an adequate set to conquer the challenge of the Sun's dynamo.

A flurry of research activity taking place throughout the 1980s, systematically studying magnetic structures with a diverse range of spatial scales, offered an enhanced picture of the Sun's magnetic evolution and extended this ``butterfly'' picture of magnetic activity. Volume 110 of the Solar Physics Journal in 1987 captures the breadth of this activity (Wilson, 1987). In concert, these primarily observational studies of ``ephemeral'' active regions, XRay (and later EUV) ``bright points'', plage, faculae, filaments, prominences, that when combined with global-scale flow patterns and characterizations of the coronal structure above the limb characterized  demonstrated a pattern that became known as the ``Extended Solar Cycle'' \-- upon realizing that the wings of the sunspot butterfly could be extended to much higher solar latitudes ($\sim55^{\circ}$ latitude) and to earlier times (almost a decade earlier) to a degree where these activity wings form a chevron\--like pattern with a strong spatio-temporal overlap in each solar hemisphere (see, e.g., Wilson et al., 1988).

A variety of observations have now asserted that the above picture of solar activity begins at higher latitudes on the Sun {\em years} before the emergence of the fresh sunspots of the coming new cycle, and they create new activity bands with the longer period of the extended solar cycle (Wilson et al., 1988). The ESCs are observed in many ways, all the way from sub-photospheric zonal flows to global-scale morphology of the corona. Altrock (1997) has observed the evidence of ESCs in the Fe XIV green line emissions in the solar corona, while Juckett (1998) has obtained evidence for a 17-year extended solar cycle in the IMF directions at 1 AU in the coronal hole variations and also in the planetary magnetospheric modulations. Robbrecht et al. (2010) have described that the extended cycle in coronal emission is not an early activity of the new solar cycle, rather it is the poleward concentration of trailing-polarity flux of the old solar cycle generated most likely by the meridional flows. Therefore, the signature of the extended solar cycle is present at diverse spatial scales at the Sun, combining various layers of its atmosphere. The observed coronal variation is linked with the presence of the torsional oscillations or zonal flow bands in the sub-surface layers of the Sun (Labonte and Howard, 1982a; Howe, 2009). 

%The most plausible mechanisms that trigger the torsional oscillations creating the ESCs are Lorentz force feedback from the magnetic cycle \citep{Sch81}, thermal feedback \citep{Spruit03}, magnetic quenching of small-scale turbulent angular-momentum transport \citep{Kru99}, modulation of angular-momentum transport by large scale meridional flows \citep{Bea13}, or the joint effect of all these physical processes.

Clearly, the ESC, magnetic activity, and sunspot cycles are intrinsically linked, but little modeling effort has been dedicated to pursuing a complete physical description of the system. In part, that is because the observational picture, and related metrics, have not been comprehensively presented for use by the modeling community.

%In this paper we briefly review the concept of an ``Extended Solar Cycle'' (ESC). In Sect.~2, we demonstrate the observational patterns of the ESC through example, while discussing some of the historical perspective. Sect.~3 describes the possible underlying mechanisms that may originate ESCs. The diagnostic capability of ESCs in understanding new cycles is described in Sect.~4. The last section discusses the importance of ESCs, as well as the future upcoming trends of ESCs using the latest ground- and space-based observations and cutting-edge models.

In the following sections we'll extend the preceding historical narrative before we discuss ESCs in contemporary observations and the possible link between the ESC and the modulation of sunspots. Finally, we'll look at the outlook of our team activity and the approach that we'll take to bring observation and model together with a goal to reduce the number of free parameters on the latter, while using the former to develop a picture of the Sun's recent climatology.

\section{ESC - A Brief Historical Perspective}
\begin{figure}[!t]
\begin{center}
\includegraphics[width=15cm]{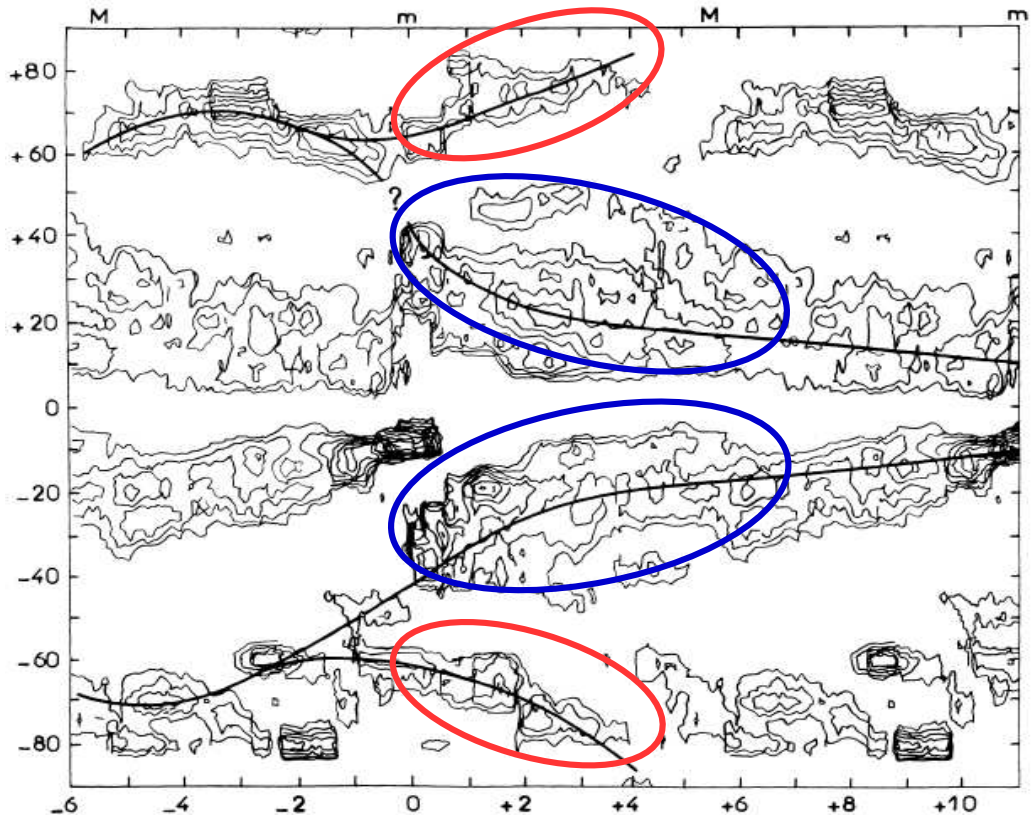}
\end{center}
\caption{The latitudinal of coronal emission and morphology obtained from Pic du Midi Observatory from 1944 to 1974 compared across cycles by aligning minima ``m'' and maxima ``M'' across the cycles in a form of superposed epoch analysis. Leroy and Noens (1983) noted that the pattern of coronal activity had an apparent (recurrent) period of about 17 years (Credit: EDP).}\label{fig:2}
\end{figure}

As mentioned above, observations of the ESC began three to four decades ago when Leroy and Noens
(1983) analyzed coronal data obtained at Pic du Midi from 1944 to 1974 and searched for the latitude variation of coronal activity, see Fig.~\ref{fig:2}. Similar work was conducted from a collection of ground-based coronagraphs and dedicated experiments identified the same patterns. These works collectively identified an underlying, global-scale, pattern of coronal activity that appeared to repeat over a period of about 17 years \-- much longer that the typical eleven-year sunspot cycle. This was consistent with an earlier work of Legrand and Simon (1981), based on the analysis of 100 years of  geomagnetic indices, which inferred the presence of two solar cycles being present on the Sun at all times, in the form of spatially overlapping cycles. In these papers, the solar cycle starts every eleventh year with a total duration of 17-18 years, and during an interval of the next 6-7 years, two consecutive cycles with different levels of activity evolved simultaneously at different latitudes of the sun. Leroy and Noens (1983) demonstrated the high-latitude (rush to the poles; red-ellipse in Fig.~\ref{fig:2}) and low-latitude (sunspot-related drift towards the equator; blue-ellipse in Fig.~\ref{fig:2}) branches of green line (Fe XIV 5303 \AA~) coronal emissions. 

Wilson et al. (1988), following on concentrated effort from co-authors over a number of years, reported on the appearance of a high-latitude accumulation of ephemeral active regions (e.g.,
Harvey and Martin, 1973; Harvey et al., 1975), the torsional oscillation (e.g., Howard and Labonte, 1980;
Labonte and Howard, 1982b), and the evolution of coronal emission (e.g., Bretz and Billings, 1959;
Altrock, 1988) as above in the diminishing phase of sunspot cycle 21. For example, the orientation of the magnetic polarities in those ephemeral active regions studied by Harvey and Martin was similar to the spots that emerged for the next sunspot cycle 22 {\em instead} of those belonging to sunspot cycle 21 \-- this striking pattern was difficult to explain at the time. {\em Concerted, these observations inferred that sunspot activity was the main phase of a more extended cycle that was triggered at higher solar latitudes prior to the maximum of the given solar cycle}. It was found to progress towards the equator during the next 18\--22 years while merging with the conventional butterfly diagram when it entered into the sunspot latitudes. Similar patterns of evolution were visible in monitoring the latitudinal progression of filaments and prominences over time (e.g., Bocchino, 1933; Hansen and Hansen, 1975) that are summarized by Cliver (2014).

\section{The Contemporary ESC}
The Solar and Heliospheric Observatory (SoHO), and following space-based observatories, supplemented the long synoptic timelines provided by ground-based synoptic observing programs. Indeed, as the ability to secure the funding of the latter has become more precarious, the space based platforms, while less flexible in terms of instrumentation, have provided a uniform, high signal-to-noise platform to augment the few (adequately funded) programs that exist on the ground. Several studies investigated the ESC. Offering a different perspective from those discussed earlier, Robbrecht et al. (2010) analyzed Fe XIV 5303\AA~green line coronal emissions using SoHO/EIT (1996\---2009) and simulated fluxes (1967\---2009), and studied the high-latitude coronal emissions associated with the coronal hole boundaries. They observed equatorward and thereafter poleward progressions of the activity bands of high-latitude coronal emissions and the associated polar crown filament following the polarity reversal. They found that strong underlying fields emerged in the minimum phase of the solar cycle by the poleward transport of the active region flux due to the surface meridional flow. These high-latitude structures were not the precursor of the new solar cycle, but they constitute a physically isolated U-shaped band moving latitudinally upward again as the active regions emerge at the mid-latitudes and reconnect with the polar coronal hole boundaries. Robbrecht et al. (2010) concluded that the extended $\approx$17 years cycle in coronal emission is {\em not} a signature of early new solar activity. Instead it was the poleward concentration of the trailing polarities of the old solar cycle caused by the meridional flow. 

Petrie et al. (2014) found gave a different explanation to the high-latitude coronal emission bands observed by Robbrecht et al. (2010), and inferred that these features may indeed have a strong connection with the main cycle under the progression, and thus the match with the  torsional oscillations (distributed around 45$^{o}$ latitude) and photospheric ephemeral bipoles having broad latitudinal distributions. 

Tappin and Altrock (2013) studied the white light coronagraph and found that similar behaviour of the global-scale corona with the historical green emission coronal observations. Altrock (2014) analyzed the ``Rush to the Poles'' for solar cycle 24 finding that the solar cycle 24 displayed an unusually intermittent rush that is only well defined in the northern hemisphere. They determined that the solar maximum in the northern hemisphere has already occurred at 2011, while it was poorly defined in the southern hemisphere.

\subsection{The ESC and the Sunspot Cycle}
The latitudinal dependence of the solar activity markers smaller in scale than sunspots (like, for example, bright points, granulation, diffuse coronal emissions, filaments/prominences, etc.) show a narrow concentration of activity that appears at higher latitudes (around 55$^{\circ}$) just after solar maximum. A portion of that concentration subsequently migrates towards the equator while another moves towards the poles. In a butterfly diagram this pattern gives the appearance of a ``bifurcation,'' or fork in activity that follows solar maximum (in each hemisphere). The former branch has been largely under-investigated with the exception of the torsional oscillation work discussed above. The latter branch has been dubbed ``the rush to the poles,'' and seems to be intimately involved in the reversal of the polar magnetic field. The equator-moving branches in each hemisphere can be traced using the same host of markers, and roughly stay equidistant to the lower latitude bands which are currently producing sunspots, moving slowly, before eventually reaching $\sim$35$^\circ$ and giving birth to the sunspots that become the next sunspot cycle. This new cycle growth appears to coincide, in very short order, with the last vestiges of the old cycle disappearing at the equator (e.g., McIntosh et al., 2014a).

\begin{figure}[!t]
\begin{center}
\includegraphics[width=15cm]{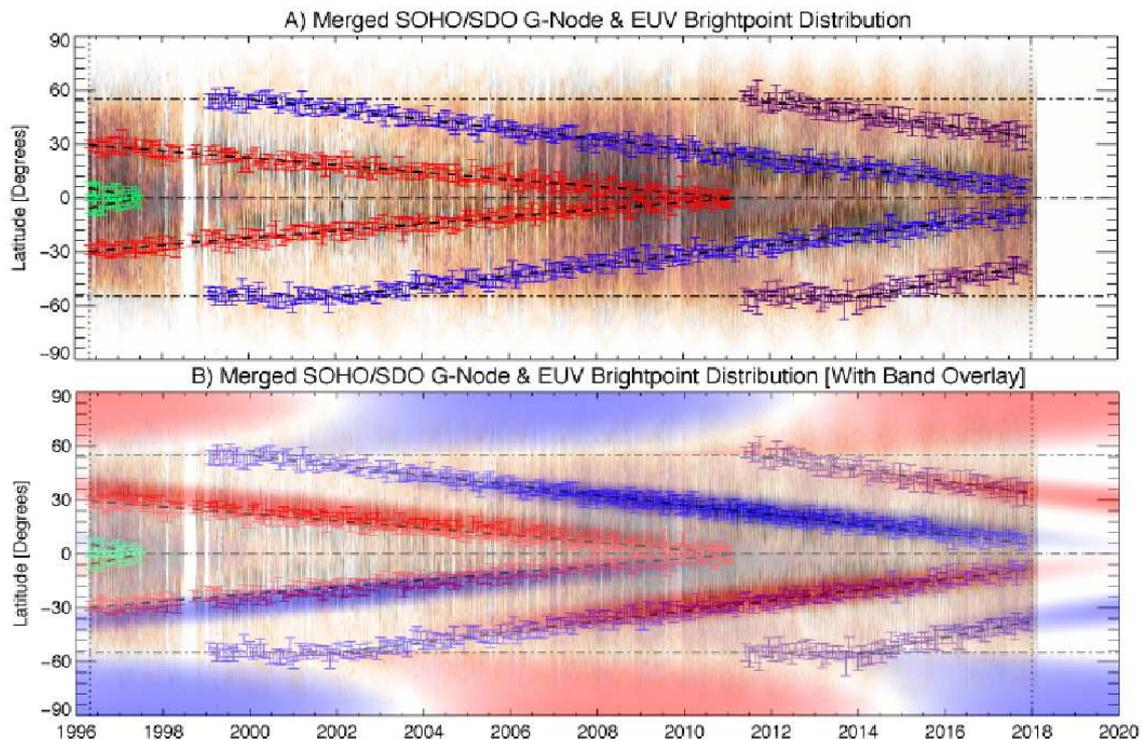}
\end{center}
\caption{Illustrating the appearance of the ESC in the contemporary era, extending the analysis of McIntosh et al. (2014a) to the present. Panel A shows the latitudinal variation of EUV bright points and their associated magnetic scale marker ``g-nodes'' McIntosh et al. (2014b) as observed respectively by SoHO and SDO. Fitting the bright point and g-node bands in latitude-time distribution allows us to track the activity bands of solar cycle 22 (green), 23 (red), 24 (blue) and 25 (purple). Note that cycle 22 terminates in 1997, cycle 23 terminates in 2011, and that we anticipate cycle 24 to end in late 2019 or early 2020. Panel B shows something called a ``band-o-gram,'' a schematic motivated by the data, to illustrate the magnetic polarities and potential interaction of the magnetic systems present.}\label{f2}
\end{figure}

Fig.~\ref{f2} provides an update to the diagnostics presented in McIntosh et al. (2014a) showing the appearance of the ESC through 2017. Extending their analysis by 4.5 years and now including the tracking of the activity bands that will give birth to the sunspots of cycle 25 \-- predicted in McIntosh et al. (2014a), and documented in McIntosh and Leamon (2017). The panels of the figure illustrate the latitudinal variation of EUV bright points and Magnetic Range of Influence (MRoI) ``g-nodes'', the magnetic features at which EUV brightpoints appear to re-occur (McIntosh et al., 2014b), as observed respectively by SoHO and SDO. The trails of bright points and g-nodes allowed the tracking of bands belonging to four sunspot cycles, 22 (green), 23 (red), 24 (blue), and 25 (purple). The lower panel shows the ``band-o-gram,'' or a data-inspired schematic of the band evolution and magnetic polarity. McIntosh et al. (2014a) were able to fit histograms to the latitudinal distributions of these features and reported that the activity bands move at speeds around 3$^{\circ}$/Yr. And that their start times, the times at which they leave 55$^{\circ}$, are asymmetric in time by as much as two years while their termination point at the equator appeared to be symmetric. It was noted by McIntosh et al. (2014a) that the departure of the bands at 55$^{\circ}$ appeared to coincide with the sunspot maximum of each hemisphere. Using some seventy years of data, McIntosh et al. (2014a) noted that the hemispheric sunspot maxima of the magnetic activity cycle (those with the same leading magnetic polarity) were approximately 22 years apart. They then proposed that the bands of cycle 25 should present themselves at 55$^{\circ}$ latitude before 2012 before leaving those latitudes offset by approximately two years \-- based on the hemispheric sunspot maxima of sunspot cycle 22 \-- as the likely launch point of the activity bands that formed sunspot cycle 23. In this updated figure, at the beginning of 2012 near the maximum of solar cycle 24, the fresh activity band appears at higher latitudes (purple) and migrates towards the solar equator and was reported in McIntosh and Leamon (2017) as the first signature of the bands that will give rise to sunspot cycle 25 in late 2019 or 2020 \-- following the expected termination of the cycle 24 bands later next year.

McIntosh et al. (2014a) was the first work to phenomenologically connect the sunspot cycle, and its temporal landmarks (minimum and maximum), with the phasing of the ESC. Noting that, in this picture, sunspot minimum is a time when there appear to be four magnetic bands within 40$^{\circ}$ of the equator and there is an appearance of mutual cancellation that is only broken (finally) by the termination of the equatorial bands.  

\section{Current Research on ESC and Its Physical Implications}
It should be noted that, as an observational induction, the peculiar behavior of sunspot progression in different latitudinal zones on the Sun was documented in the 19th century by keen observers such as Carrington. This early work laid the foundation for the discovery of the ESC when those results were summarized later (e.g., Clerke, 1903). However we do note that, despite the considerable observational effort spent, and the connection of the various features (from small to global scales) with an evolutionary path that is eventually populated by sunspots, very little effort has been spent on addressing the nature of the ESC in relation to the Sun's dynamo \-- our focused team hopes to do that. In the remainder of the paper we'll discuss some of the questions and approaches required to close the (considerable) gap between observation and model.

While these efforts using space-borne EUV imagers, coronagraphs, and first-generation ground-based observations tried to shed new light on the behavior of extended solar cycles, dramatic changes have occurred in past ten years with the advent of new-generation high-resolution observatories, and associated more advanced computational and image-processing techniques. The availability of a wealth of high-resolution observations from ground and space about small-scale features (e.g., EUV bright points, g-nodes, etc.) have enabled the study of the properties of ESCs in greater detail, their possible linkage with the sub-surface dynamics, and inherent physical implications. Apart from the classical solar cycle markers, e.g., sunspots, plage, filaments, etc., in the last few years, the new observational markers have been discovered and utilized to understand the behavior of solar variability, e.g., EUV bright points, g-nodes associated with the evolution of the rotationally driven giant convective scales (McIntosh et al.,
2014c). In concert, all of these features provide the opportunity to track the progression of solar activity bands beyond the traditional eleven-year solar cycle.

The perspective of our team is first to develop and then deploy a consistent approach to the analysis of the space and ground-based data available to us. All in all we have data that spans back from the present day to the dawn of H$_{\alpha}$ photography in the late 1870s. From that we will be able to asses key features such as band onset times, times of bifurcation, latitudinal migration speeds, termination points, etc that can then be used to develop an enhanced set of boundary conditions for model developers. Stand-alone, these numbers over some 140 years, a dozen sunspot cycles, form an important benchmark, and their averages form a climatology can become a consensus set of parameters describing the ESC. The task for the modeling part of our team is then to try and incorporate these markers, timescales, etc into the framework of the contemporary models. Which of the contemporary models can and which cannot describe the ESC? Can they help us understand why the region around 55$^{\circ}$ is so important (apparently) to the evolution observed? Can the incorporation of the ESC information help to improve the numerical forecasting of sunspot cycle timing and amplitude? 

%The novelty of these current trends is the forecast capability of the solar activity cycles using measured progression of the magnetic activity bands of the particular cycle, projected onset time of new activity bands appearing at the higher latitudes, and average migration speed towards the equator \citep{Scott14b}. For example, Fig.~3 shows the updated version of the butterfly diagram (up to 2018 in the current solar cycle) of EUV bright-points (BPs) as observed initially by \citet{Scott17} using SDO AIA 193 \AA~observations. Extended solar cycles of cycle 24 and upcoming cycle 25 have been respectively shown by white and pink lines. Magnetic activity bands that will trigger solar cycle 25 are already visible at higher latitudes before the termination of the equatorward drift of the activity bands of the previous solar cycle 24 somewhere in 2019. 

\section{Concluding Remarks}
The extended solar cycle has been readily observed for many decades. Our team will bring together the various observational datasets and apply a consistent approach to their analysis. We will then publish a consensus climatology for the ESC over the 140 years of observed evolution. The (considerable) challenge is then to understand, via modeling activities, how the ESC helps to better understand the interior of the Sun and constrain models of the solar dynamo. Maybe, following that last step, we can begin to explore how the activity of stars with different luminosities, interior structure, and rotation rates may be understood using our new solar basis as a ``Rosetta Stone.''

\section*{Conflict of Interest Statement}
The authors declare that the research was conducted in the absence of any commercial or financial relationships that could be construed as a potential conflict of interest.

\section*{Author Contributions}
All authors have contributed in an equal manner. All the member co-authors of Indo-US (IUSSTF) Joint Networked R\&D Center team have worked, read, and edit the manuscript.

\section*{Funding}
Indo-US (IUSSTF) Joint Networked R\&D Center (Ref: IUSSTF-JC-011-2016)

\section*{Acknowledgments}

All author acknowledge Indo-US (IUSSTF) Joint Networked R\&D Center (Ref: IUSSTF-JC-011-2016).

%\bibliographystyle{frontiersinSCNS_ENG_HUMS} % for Science, Engineering and Humanities and Social Sciences articles, for Humanities and Social Sciences articles please include page numbers in the in-text citations
%\bibliographystyle{frontiersinHLTH&FPHY} % for Health, Physics and Mathematics articles
%\bibliography{test}

\section*{References}
Altrock, R. C. (1988). Variation of solar coronal Fe XIV 5303 A emission during Solar Cycle 21. In Solar and Stellar Coronal Structure and Dynamics, ed. R. C. Altrock. 414–420\\
Altrock, R. C. (1997). An extended Solar CYCLE as Observed in fe XIV. Sol. Phys. 170, 411–423. doi:10.1023/A:1004958900477 \\
Altrock, R. C. (2014). Forecasting the Maxima of Solar Cycle 24 with Coronal Fe xiv Emission. Sol. Phys. 289, 623–629. doi:10.1007/s11207-012-0216-1 \\
Babcock, H. W. (1961). The Topology of the Sun’s Magnetic Field and the 22-YEAR Cycle. Astrophys. J. 133, 572. doi:10.1086/147060 \\
Bocchino, G. (1933). Migrazione delle protuberanze durante il ciclo undecennale dell’attività solare. Osservazioni e memorie dell’Osservatorio astrofisico di Arcetri 51, 5–47 \\
Bretz, M. C. and Billings, D. E. (1959). Analysis of Emission Corona 1942-1955 from Climax Spectrograms. Astrophys. J. 129, 134. doi:10.1086/146601 \\
Charbonneau, P. (2010). Dynamo Models of the Solar Cycle. Living Reviews in Solar Physics 7, 3.doi:10.12942/lrsp-2010-3 \\
Choudhuri, A. R., Chatterjee, P., and Jiang, J. (2007). Predicting Solar Cycle 24 With a Solar Dynamo Model. Physical Review Letters 98, 131103. doi:10.1103/PhysRevLett.98.131103 \\
Choudhuri, A. R., Schussler, M., and Dikpati, M. (1995). The solar dynamo with meridional circulation. Astron. Astrophys. 303, L29 \\
Clerke, A. M. (1903). Problems in Astrophysics Cliver, E. W. (2014). The Extended Cycle of Solar Activity and the Sun’s 22-Year Magnetic Cycle. Sp. Sci. Rev. 186, 169–189. doi:10.1007/s11214-014-0093-z \\
Dikpati, M. and Gilman, P. A. (2006). Simulating and Predicting Solar Cycles Using a Flux-Transport Dynamo. Astrophys. J. 649, 498–514. doi:10.1086/506314 \\
Fan, Y. (2009). Magnetic Fields in the Solar Convection Zone. Living Reviews in Solar Physics 6, 4. doi:10.12942/lrsp-2009-4 \\
Hale, G. E. (1908). On the Probable Existence of a Magnetic Field in Sun-Spots. Astrophys. J. 28, 315. doi:10.1086/141602 \\
Hale, G. E., Ellerman, F., Nicholson, S. B., and Joy, A. H. (1919). The Magnetic Polarity of Sun-Spots. Astrophys. J. 49, 153. doi:10.1086/142452 \\
Hale, G. E. and Nicholson, S. B. (1925). The Law of Sun-Spot Polarity. Astrophys. J. 62, 270. doi:10. 1086/142933 \\
Hansen, R. and Hansen, S. (1975). Global distribution of filaments during solar cycle No. 20. Sol. Phys. 44, 225–230. doi:10.1007/BF00156857 \\
Harvey, K. L., Harvey, J. W., and Martin, S. F. (1975). Ephemeral active regions in 1970 and 1973. Solar Phys. 40, 87–102. doi:10.1007/BF00183154 \\
Harvey, K. L. and Martin, S. F. (1973). Ephemeral Active Regions. Solar Phys. 32, 389–402. doi:10.1007/BF00154951 \\
Hathaway, D. H. (2010). The Solar Cycle. Living Reviews in Solar Physics 7, 1. doi:10.12942/lrsp-2010-1 \\
Howard, R. and Labonte, B. J. (1980). The sun is observed to be a torsional oscillator with a period of 11 years. Astrophys. J. Lett. 239, L33–L36. doi:10.1086/183286 \\
Howe, R. (2009). Solar Interior Rotation and its Variation. Living Reviews in Solar Physics 6, 1. doi:10.12942/lrsp-2009-1 \\
Jiang, J., Chatterjee, P., and Choudhuri, A. R. (2007). Solar activity forecast with a dynamo model. Monthly Notices of Royal Astron. Soc. 381, 1527–1542. doi:10.1111/j.1365-2966.2007.12267.x \\
Juckett, D. A. (1998). Evidence for a 17-year cycle in the IMF directions at 1 AU, in coronal hole variations, and in planetary magnetospheric modulations. Sol. Phys. 183, 201–224. doi:10.1023/A:
1005075703810 \\
Labonte, B. J. and Howard, R. (1982a). Torsional waves on the sun and the activity cycle. Sol. Phys. 75, 161–178. doi:10.1007/BF00153469 \\
Labonte, B. J. and Howard, R. (1982b). Torsional waves on the sun and the activity cycle. Sol. Phys. 75, 161–178. doi:10.1007/BF00153469 \\
Lanza, A. F. (2010). Stellar magnetic cycles. In Solar and Stellar Variability: Impact on Earth and Planets, eds. A. G. Kosovichev, A. H. Andrei, and J.-P. Rozelot. vol. 264 of IAU Symposium, 120–129.
doi:10.1017/S1743921309992523 \\
Legrand, J. P. and Simon, P. A. (1981). Ten cycles of solar and geomagnetic activity. Sol. Phys. 70, 173–195. doi:10.1007/BF00154399 \\
Leighton, R. B. (1969). A Magneto-Kinematic Model of the Solar Cycle. Astrophys. J. 156, 1. doi:10.1086/149943 \\
Leroy, J.-L. and Noens, J.-C. (1983). Does the solar activity cycle extend over more than an 11-year period? Astron. Astrophys. 120, L1 \\
Maunder, E. W. (1904). Note on the Distribution of Sun-spots in Heliographic Latitude, 1874-1902. Monthly Notices of Royal Astr. Soc. 64, 747–761. doi:10.1093/mnras/64.8.747 \\
McIntosh, S. W. and Leamon, R. J. (2017). Deciphering Solar Magnetic Activity: Spotting Solar Cycle 25. Frontiers in Astronomy and Space Sciences 4, 4. doi:10.3389/fspas.2017.00004 \\
McIntosh, S. W., Leamon, R. J., Gurman, J. B., Olive, J.-P., Cirtain, J. W., Hathaway, D. H., et al. (2013). Hemispheric Asymmetries of Solar Photospheric Magnetism: Radiative, Particulate, and Heliospheric Impacts. Astrophys. J. 765, 146. doi:10.1088/0004-637X/765/2/146 \\
McIntosh, S. W., Wang, X., Leamon, R. J., Davey, A. R., Howe, R., Krista, L. D., et al. (2014a). Deciphering Solar Magnetic Activity. I. On the Relationship between the Sunspot Cycle and the
Evolution of Small Magnetic Features. Astrophys. J. 792, 12. doi:10.1088/0004-637X/792/1/12 \\
McIntosh, S. W., Wang, X., Leamon, R. J., and Scherrer, P. H. (2014b). Identifying Potential Markers of the Sun’s Giant Convective Scale. Astrophys. J. Lett. 784, L32. doi:10.1088/2041-8205/784/2/L32 \\
McIntosh, S. W., Wang, X., Leamon, R. J., and Scherrer, P. H. (2014c). Identifying Potential Markers of the Sun’s Giant Convective Scale. Astrophys. J. 784, L32. doi:10.1088/2041-8205/784/2/L32 \\
Parker, E. N. (1955). Hydromagnetic Dynamo Models. Astrophys. J. 122, 293. doi:10.1086/146087 \\
Petrie, G. J. D., Petrovay, K., and Schatten, K. (2014). Solar Polar Fields and the 22-Year Activity Cycle:Observations and Models. Space Sci. Rev. 186, 325–357. doi:10.1007/s11214-014-0064-4 \\
Robbrecht, E., Wang, Y.-M., Sheeley, N. R., Jr., and Rich, N. B. (2010). On the Extended Solar Cycle in Coronal Emission. Astrophys. J. 716, 693-700. doi:10.1088/0004-637X/716/1/693 \\
Svalgaard, L. (2013). Solar activity - past, present, future. Journal of Space Weather and Space Climate 3, A24. doi:10.1051/swsc/2013046 \\
Tappin, S. J. and Altrock, R. C. (2013). The Extended Solar Cycle Tracked High into the Corona. Sol. Phys. 282, 249–261. doi:10.1007/s11207-012-0133-3 \\
Usoskin, I. G. (2017). A history of solar activity over millennia. Living Reviews in Solar Physics 14, 3. doi:10.1007/s41116-017-0006-9 \\
Vizoso, G. and Ballester, J. L. (1990). The north-south asymmetry of sunspots. Astron. Astrophys. 229, 540–546 \\
Wang, Y.-M. and Sheeley, N. R., Jr. (1991). Magnetic flux transport and the sun’s dipole moment - New twists to the Babcock-Leighton model. Astrophys. J. 375, 761–770. doi:10.1086/170240 \\
Wilson, P. R. (1987). Solar Cycle Workshop. Proceedings of the 2nd meeting, held at Stanford Sierra Lodge, Lake Tahoe, California, 10 - 14 May 1987. \\
Wilson, P. R., Altrock, R. C., Harvey, K. L., Martin, S. F., and Snodgrass, H. B. (1988). The extended solar activity cycle. Nature 333, 748–750. doi:10.1038/333748a0 \\
Yeates, A. R., Nandy, D., and Mackay, D. H. (2008). Exploring the Physical Basis of Solar Cycle Predictions: Flux Transport Dynamics and Persistence of Memory in Advection- versus Diffusion-
dominated Solar Convection Zones. Astrophys. J. 673, 544-556. doi:10.1086/524352 \\
Yoshimura, H. (1975). Solar-cycle dynamo wave propagation. Astrophys. J. 201, 740–748. doi:10.1086/153940

%%% If you are submitting a figure with subfigures please combine these into one image file with part labels integrated.
%%% If you don't add the figures in the LaTeX files, please upload them when submitting the article.
%%% Frontiers will add the figures at the end of the provisional pdf automatically
%%% The use of LaTeX coding to draw Diagrams/Figures/Structures should be avoided. They should be external callouts including graphics.
\end{document}